\documentclass[amsmath]{revtex4-1}
\usepackage[utf8]{inputenc}
\usepackage{amsfonts}
\usepackage{amsmath}
\usepackage{amssymb}
\usepackage{graphicx}%
\newcommand{\be}{\begin{equation}}
\newcommand{\ee}{\end{equation}}
\newcommand{\bea}{\begin{eqnarray}}
\newcommand{\eea}{\end{eqnarray}}

\begin{document}
\title{Exchange contribution to the variance of the excitation energy of the electron shell \\
of the daughter atom in double--$\beta$ decay}

\author{K. S. Tyrin}~
\email{E-mail address: Tyrin\_KS@nrcki.ru}
\affiliation{National Research Centre ``Kurchatov Institute'', 
               123182 Moscow, Russia}
\author{M. I. Krivoruchenko}
\email{E-mail address: mikhail.krivoruchenko@itep.ru}
\affiliation{National Research Centre ``Kurchatov Institute'', 
               123182 Moscow, Russia}
               
\begin{abstract}
The excitation of electron shell of a daughter atom in a neutrinoless double $\beta$--decay causes
change in shape of the total energy peak of $\beta$ electrons at the end of the energy spectrum. 
The main parameters of the modified distribution are the average energy and variance of the excitation energy of the electron shell. We derive an expression for the variance taking into account exchange effects and make numerical estimates
of the average excitation energy and variance based on the non-relativistic Roothaan--Hartree--Fock method and the relativistic
Dirac--Hartree--Fock method implemented in 
the General Relativistic Atomic Structure Package (G\textsc{rasp}2018). Estimates
are made for eleven isotopes, two--neutrino double--$\beta$ decay of which is observed
experimentally. The results are determined by the first two negative moments
of the electron radii in the parent atom. The values obtained for change in the peak shape 
can be used to parameterize the energy distribution of $\beta$ electrons, taking into account
the excitation of the electron shell of the daughter atom.
\end{abstract}

\maketitle
 
\section{Introduction}

The study of $\beta$ processes sensitive to mixing, mass, and nature of neutrinos (Dirac/Majorana) is a promising way in the search for extensions of the Standard Model (SM). The neutrinoless double--$\beta$ ($0\nu2\beta$) decay is important for testing the conservation of the total lepton number and the nature of neutrinos (for a review see \cite{Bilenky:2018}). Observing $0\nu2\beta$ decay indicates the presence of a Majorana neutrino mass, according to the "black box theorem"~\cite{Schechter:1982,Hirsch:2006}. In the quark sector of the SM, proton decay and neutron-antineutron oscillations play a similar role in testing the total baryon number conservation. The limitation of the period of neutron-antineutron oscillations restricts the neutron Majorana mass to $\Delta m_n < 0.8\times 10^{-33} m_n$, where $m_n = 939.57$ MeV/c$^2$ is the neutron Dirac mass \cite{Blaum:2020}. The Majorana neutrino mass is limited by a fraction of electronvolts \cite{Simkovic:2021}.

In $0\nu2\beta$ decay, the decay energy $Q$ is distributed between two $\beta$ electrons. Thus, their total energy equals the value of $Q$. The occurrence of a peak at energy $Q$ against the background of a continuous distribution of electron energy in a two--neutrino double--$\beta$ decay ($2\nu2\beta$) is the signature of $0\nu2\beta$--decay. The collaborations CUORE \cite{Adams:2020,Adams:2022}, EXO \cite{Anton:2019}, KamLAND-Zen \cite{Abe:2023} and LEGEND \cite{Legend} are searching for $0\nu2\beta$--decay to clarify the Dirac/Majorana nature of neutrinos and estimate their Majorana mass. Multielectron modes can be seen while examining the $\beta$ electron spectrum of the SuperNEMO collaboration \cite{Arnold:2014,Arnold:2015}.

The limited energy resolution of detector, together with the nuclear recoil and excitation of the parent atom's electron shell, modify the energy spectrum of $\beta$ electrons. In $\beta$ processes, a change in the charge of the nucleus acts on the electron shell as a sudden perturbation, forcing the electrons of a daughter ion to move to excited levels (shake-up) or ionization states of a continuous spectrum (shake-off) with a certain probability. The energy probability distribution splits into two terms: the first in which all electrons preserve quantum numbers and the reaction energy is not modified, and the second in which excitation and ionization occur.
In the second situation, the effective decay energy decreases, which affects the spectrum of $\beta$ electrons around the boundary value, sensitive to the Majorana neutrino mass.

Modeling the probability distribution of $\beta$ electrons in $0\nu2\beta$--decay requires estimates of the average excitation energy and variance of the daughter ion's electron shell. The excitation effects are discussed in \cite{Krivoruchenko:2023a,Krivoruchenko:2023b}. The average excitation energy and variance of excitation energy of the electron shell are estimated, taking exchange effects into account. In Section II.B, we provide derivation of the formulas used
in \cite{Krivoruchenko:2023b} to calculate exchange effects. The variance is estimated with the use of the matrix elements of $r^{-1}$ and $r^{-2}$, which account for exchange effects. Section III presents the matrix elements of $r^{-1}$ and $r^{-2}$ for eleven isotopes, whose double--$\beta$ decays are observed experimentally. The non-relativistic Roothaan--Hartree--Fock (RHF) method \cite{Clementi:1974} and the software package G\textsc{rasp}2018 \cite{Dyall:1989, Fischer:2019, Grant:2008}, based on the relativistic Dirac--Hartree--Fock (DHF) method, are employed for calculations.

\section{Excitation energy of electron shell and variance}
\renewcommand{\theequation}{II.\arabic{equation}}
\setcounter{equation}{0}


The paper discusses $0\nu2\beta$ decay with decay energy $Q\gtrsim mc^2$, where $m$ is the electron mass and $c$ is the speed of light. This criterion is fulfilled for all eleven isotopes, 
whose $2\nu2\beta$--decays were observed (see Table \ref{table5}). The condition $Q\gtrsim mc^2$ indicates that the velocity of $\beta$ electrons approaches the speed of light and exceeds the typical velocity $\sim\alpha Z^{1/3} c $ of electrons in multielectron atoms.


\begin{table}[t]
\addtolength{\tabcolsep}{0pt} 
\caption{
Variance of the excitation energy of the electron shells of the daughter atom for eleven isotopes, the $0\nu2\beta$--decays of which are observed experimentally. The second column contains the values of the mass difference, $Q$, of the neutral atoms involved in the decay. The fourth column presents the average excitation energies calculated by the DHF method using the G\textsc{rasp}2018 package.
The fifth and sixth columns show the exchange contributions to the variance and the total variance 
obtained in the framework of the non-relativistic RHF method. 
The last two columns report the exchange contributions and the total variance obtained using the relativistic DHF method implemented into the G\textsc{rasp}2018.
\label{table5}
} 
\renewcommand{\arraystretch}{1.2} {\scriptsize \centering\vspace{2mm} 
\begin{tabular}{|rrr@{\hspace{2mm}}lc@{\hspace{2mm}}ccccc|}
\hline\hline
\multicolumn{3}{|c}{Process} & $%
\begin{array}{c}
Q \\ 
\mathrm{[keV]}%
\end{array}%
$ & Ref. & $%
\begin{array}{c}
\mathcal{C}_{\mathrm{DHF}} \\ 
\mathrm{[eV]}%
\end{array}%
$ & $%
\begin{array}{c}
\Delta \mathcal{D}_{\mathrm{RHF}}^{1/2} \\ 
\mathrm{[keV]}%
\end{array}%
$ & $%
\begin{array}{c}
\mathcal{D}_{\mathrm{RHF}}^{1/2} \\ 
\mathrm{[keV]}
\end{array}%
$ & $%
\begin{array}{c}
\Delta \mathcal{D}_{\mathrm{DHF}}^{1/2} \\ 
\mathrm{[keV]}%
\end{array}%
$ & $%
\begin{array}{c}
\mathcal{D}_{\mathrm{DHF}}^{1/2} \\ 
\mathrm{[keV]}
\end{array}%
$ \\ \hline
$_{\;\,20}^{\;\,48}\mathrm{Ca}$ & $\rightarrow $ & $_{\;\,22}^{\;\,48}\mathrm{Ti}$ & 4267.98(32) & \cite{Kwiatkowski:2014} &283& -0.05 & 1.61 & -0.05 & 1.64\\ 
$_{\;\,32}^{\;\,76}\mathrm{Ge}$ & $\rightarrow $ & $_{\;\,34}^{\;\,76}\mathrm{Se}$ & 2039.006(50) & \cite{Suhonen:2007} &365& -0.10 & 2.62 & -0.11 & 2.77 \\ 
&  &  & 2039.061(7) & \cite{GERDA:2017} &  &  & && \\ 
$_{\;\,34}^{\;\,82}\mathrm{Se}\,$ & $\rightarrow $ & $_{\;\,36}^{\;\,82}\mathrm{Kr}$ & 2997.9(3) & \cite{Lincoln:2013} &375& -0.11 & 2.79 & -0.12 & 2.97 \\ 
$_{\;\,40}^{\;\,96}\mathrm{Zr}\,$ & $\rightarrow $ & $_{\;\,42}^{\;\,96}\mathrm{Mo}$ & 3356.097(86) & \cite{Alanssari:2016} &403& -0.15 & 3.29 & -0.15 & 3.60 \\ 
$_{\;\,42}^{100}\mathrm{Mo}$ & $\rightarrow $ & $_{\;\,44}^{100}\mathrm{Ru}$ & 3034.40(17) & \cite{Rahaman:2008} &416& -0.16 & 3.46 & -0.17 & 3.83 \\ 
$_{\;\,48}^{116}\mathrm{Cd}$ & $\rightarrow $ & $_{\;\,50}^{116}\mathrm{Sn}$ & 2813.50(13) & \cite{Rahaman:2011} &448& -0.20 & 3.97 & -0.21 & 4.54 \\ 
$_{\;\,52}^{128}\mathrm{Te}\,$ & $\rightarrow $ & $_{\;\,54}^{128}\mathrm{Xe} $ & 865.87(131) & \cite{Scielzo:2009} &468& -0.21 & 4.32 & -0.23 & 5.05 \\ 
$_{\;\,52}^{130}\mathrm{Te}\,$ & $\rightarrow $ & $_{\;\,54}^{130}\mathrm{Xe} $ & 2526.97(23) & \cite{Rahaman:2011} &468& -0.21 & 4.32 & -0.23 & 5.05 \\ 
$_{\;\,54}^{136}\mathrm{Xe}$ & $\rightarrow $ & $_{\;\,56}^{136}\mathrm{Ba}$ & 2457.83(37) & \cite{Redshaw:2007} &475& -0.22 & 4.49 & -0.25 & 5.32 \\ 
$_{\;\,60}^{150}\mathrm{Nd}$ & $\rightarrow $ & $_{\;\,62}^{150}\mathrm{Sm}$ & 3371.38(20) & \cite{Kolhinen:2010} &515&  & & -0.29 & 6.20 \\ 
$_{\;\,92}^{238}\mathrm{U}\;\;$ & $\rightarrow $ & $_{\;\,94}^{238}\mathrm{Pu}$ & 1437.3 & \cite{Firestone:1996} &817&  & & -0.63 & 13.95  \\ 
\hline\hline
\end{tabular}%
} 
\end{table}

In $\beta$--decay problems, an approximation of a sudden perturbation gives adequate precision. For an accurate approximation, the flight duration $\tau_{f}$  of $\beta$ electron through the atomic shell should be smaller than the period  $\tau_{o}$ of electron's revolution around the nucleus (see, e.g., \cite{Landau.Lifschitz:1987}).  The energy of $\beta$ electrons follows a continuous distribution. When one of the $\beta$ electrons carries energy alongside the phase space volume boundary, the velocity of the second electron is minimal. In such a situation, the approach of a sudden perturbation of the atomic shell is inequitable. 
The  usefulness of approximation is confined to the middle region of the energy spectrum, where $\beta$ electron energies are comparable and neutrinos have low energy. In $0\nu2\beta$ decay, only
electrons carry away the decay energy, and the concept of a sudden perturbation has a broader range of validity.

In what follows, the atomic system of units $\hbar = m = e^2 = 1$, $c = 137$ is employed. The $\beta$ electron flight duration through the shell, $\tau_{f}$, can be estimated as the ratio of the K-orbit radius 
$a_K \sim 1/Z$ to the velocity of the $\beta$ electron $pc^2/E\sim c$ for $Q \gtrsim c^2$. The time the electron takes to rotate around the nucleus is $\tau _{o} 
\sim 2\pi / Z^2$. The perturbation is sudden if the electron on the shell does not have enough time to perform one circle around the nucleus during the flight of the $\beta$ electron. The related instantaneous parameter is
\begin{equation}
\xi _{\mathrm{inst}} = \frac{\tau _{f}}{\tau _{o}} \sim \frac{Z} {2\pi c}.
\end{equation}
For a germanium atom of $Z=32$, $\xi_{\mathrm{inst}} \sim 1/25$. For outer orbits where the shielded charge of the nucleus is of the order of one, a substitution of $Z \to Z_{\mathrm{eff}} \sim 1$ should be made, resulting in an additional decline in $\xi_{\mathrm{inst}}$. Thus, for $Q\gtrsim c^2$, the concept of suddenness of the perturbation is quite realistic.

\subsection{Multielectron atoms}

In relativistic theory, the Hamiltonian of $N$ electrons in an atom with a
nuclear charge $Z$ is written as follows \cite{Grant:2008}
\begin{equation}
\hat{H}_{Z,N}=\sum_{i=1}^{N}\left( \mbox {\boldmath$\alpha$}_{i}\mathbf{p}%
_{i}c +\beta _{i}c^2 -\frac{Z}{|\mathbf{x}_{i}|}\right) +\sum_{i<j}^{N}V_{\mathrm{%
CG/CB}}(\mathbf{x}_{i}-\mathbf{x}_{j}),  \label{hamiltonian}
\end{equation}%
where atomic units are used, 
\textbf{$\mbox {\boldmath$\alpha$}$} and $\beta $ are the Dirac gamma
matrices, $V_{\mathrm{CG/CB}}(\mathbf{x}_{i}-%
\mathbf{x}_{j})$ is the gauge-dependent lowest-order relativistic interaction potential
of two electrons. Lorenz and Coulomb gauges are usually used to model
electromagnetic interaction of electrons in multielectron atoms. 

The Coulomb-Gaunt potential refers to the interaction potential of electrons
in the Lorenz gauge \cite{Gaunt1929}: 
\begin{equation} \label{CG}
V_{\mathrm{CG}}(\mathbf{x}_{i}-\mathbf{x}_{j})=\frac{1}{|\mathbf{x}_{i}-%
\mathbf{x}_{j}|}\left( 1 - \frac{\mbox{\boldmath$\alpha$}_{i}\cdot %
\mbox{\boldmath$\alpha$}_{j}}{c^2}\right) .
\end{equation}
The matrices $\mbox {\boldmath$\alpha$}_{i}$ can be interpreted as the electron
velocity operators divided by the speed of light $c$. The potential $V_{%
\mathrm{CG}}(\mathbf{x}_{i}-\mathbf{x}_{j})$ describes the lowest-order photon-exchange static
interaction of two electrons. The second term is responsible for the energy
of magnetostatic interaction; it is of the same order $\sim v^{2}/c^{2}$ as
the retardation correction to the one-photon exchange. For the innermost shells $%
v\sim 1/r \sim Z$, where $r=|\mathbf{x}|$. 

The Coulomb-Breit static potential can be obtained using the Coulomb gauge
\cite{Breit1929}: 
\begin{equation} \label{CB}
V_{\mathrm{CB}}(\mathbf{x}_{i}-\mathbf{x}_{j})=\frac{1}{|\mathbf{x}_{i}-%
\mathbf{x}_{j}|}\left( 1-\frac{\mbox{\boldmath$\alpha$}_{i}\cdot %
\mbox{\boldmath$\alpha$}_{j}+\mbox{\boldmath$\alpha$}_{i}\cdot \mathbf{n}%
\mbox{\boldmath$\alpha$}_{j}\cdot \mathbf{n}}{2c^2}\right) ,
\end{equation}%
where $\mathbf{n}=\mathbf{(\mathbf{x}_{i}-\mathbf{x}_{j})}/|\mathbf{x}_{i}-%
\mathbf{x}_{j}|$. $V_{\mathrm{CB}}(\mathbf{x}_{i}-\mathbf{x}_{j})$ is
precise to order $1/c^{2}$, with the retardation modifications appearing
at order $1/c^{4}$. The equivalence of potentials (\ref{CG}) and (\ref{CB}) 
in the order $1/c^2$ is demonstrated in Ref.~\cite{Blaum:2020}.

These potentials are used to generate initial approximations for a variety
of atomic physics problems. The non-covariant Coulomb gauge is appropriate
for directly solving the constraint field equations, allowing a transition
from classical electrodynamics to quantum electrodynamics (QED) via canonical quantization. The
equivalence of the covariant Lorenz gauge and the non-covariant Coulomb
gauge implies relativistic covariance of QED. Quantum electrodynamics of
multielectron atoms is a gauge invariant theory in all
orders of perturbation theory in relation to atomic energy eigenvalues and
electromagnetic form factors \cite{Krivoruchenko:2023}. 

The ground state of $N$ electrons in an atom with a nuclear charge $Z$ is
represented by $|Z,N\rangle $. By the definition 
\begin{equation}
\hat{H}_{Z,N}|Z,N\rangle =E_{Z,N}|Z,N\rangle ,  \label{eigen}
\end{equation}%
where $E_{Z,N}$ is the total energy of electron shell. The energy spectrum
has discrete and continuous parts. The discrete energy eigenstates are normalized
by one: 
\begin{equation}
\langle Z,N|Z,N\rangle =1,  \label{norm}
\end{equation}%
while states with different energy eigenvalues are orthogonal. 

\subsection{Average excitation energy}

Double--$\beta$ decay increases the nuclear charge by two units. The
Coulomb addition to the Hamiltonian causes a sudden perturbation,
transforming the ground state $|Z,Z\rangle$ of the initial Hamiltonian $\hat{%
H}_{Z,Z}$ into a superposition of states with definite energy of the
final Hamiltonian: 
\begin{equation}  \label{decompose}
\hat{H}_{Z+2,Z} = \hat{H}_{Z,Z} -2 \sum_{i=1}^{Z}\frac{1}{{r}_{i}},
\end{equation}
where $r_i = |\mathbf{x}_{i}|$. The average energy of $Z$ electrons of the parent atom in the Coulomb field of the daughter
nucleus with a charge of $Z + 2$ is equal to 
\begin{eqnarray}
\langle Z,Z|\hat{H}_{Z+2,Z}|Z,Z\rangle = E_{Z,Z} + 2 E_{Z,Z}^{\mathrm{C}}/Z,
\label{Estar}
\end{eqnarray}
where $E_{Z,Z}^{\mathrm{C}}$ is the average energy of the Coulomb
interaction of electrons of the parent atom with the nucleus: 
\begin{eqnarray}  \label{Coulomb}
E_{Z,Z}^{\mathrm{C}} = - \langle Z,Z| \sum_{i=1}^{Z} \frac{Z}{r_i}%
|Z,Z\rangle .
\end{eqnarray}
The average value of the excitation energy of the electron shell of the
daughter ion is determined by \cite%
{Krivoruchenko:2023a,Krivoruchenko:2023b} 
\begin{equation}  \label{C}
\mathcal{C} = E_{Z,Z} + 2 E_{Z,Z}^{\mathrm{C}}/Z - E_{Z+2,Z}.
\end{equation}
Binding energies of neutral atoms are tabulated in Refs.~\cite{Lu:1971,
Desclaux:1973, Clementi:1974, HUANG:1976} and can be calculated with the use of
programs designed to model structure of electron shells (see, e.g., \cite{Dyall:1989,Fischer:2019}). The Coulomb interaction 
energy $E_{Z,Z}^{\mathrm{C}}$ is tabulated in Ref. \cite{HUANG:1976}. 

\subsection{Variance of excitation energy taking into account exchange
effects}

In simple scenarios, the probability distribution functions can be characterized
by two parameters, for which it is enough to know only the first two
moments. The average of the excitation energy is used as the first, and the
variance of excitation energy,
\begin{equation*}
\mathcal{D}=\langle Z,Z|\hat{H}_{Z+2}^{2}|Z,Z\rangle -\langle Z,Z|\hat{H}%
_{Z+2}|Z,Z\rangle ^{2},
\end{equation*}%
is used as the second. Taking into account Eqs.~(\ref{eigen}) and (\ref{decompose}) we have 
\begin{equation}  \label{211}
\frac{1}{4}\mathcal{D}=\langle Z,Z|\sum_{i=1}^{Z}\sum_{j=1}^{Z}\frac{1}{r_{i}%
}\frac{1}{r_{j}}|Z,Z\rangle -\langle Z,Z|\sum_{i=1}^{Z}\frac{1}{r_{i}}%
|Z,Z\rangle ^{2}.
\end{equation}%
Finding the average values of $1/r_{i}$ and $1/r_{i}1/r_{j}$ with the help of the wave
function of the parent atom  allows to calculate the variance.

It should be noted that Eqs.~(\ref{C}) and (\ref{211}) are valid for any form 
of the two-electron interaction potential $V_{\mathrm{%
CG/CB}}(\mathbf{x}_{i}-\mathbf{x}_{j})$. When deriving these equations, only 
the property
\[
\frac{\partial }{\partial Z}\hat{H}_{Z,N}=-\sum_{i=1}^{N}\frac{1}{|\mathbf{x}%
_{i}|}
\]
is used. 

Ignoring the exchange contribution, factorization holds for $i\neq j$ 
\begin{equation}
\langle Z,Z| \frac{1}{{r}_i {r}_j}|Z,Z\rangle \approx \langle Z,Z| \frac{1}{{%
r}_i}|Z,Z\rangle \langle Z,Z| \frac{1}{{r}_j}|Z,Z\rangle.
\end{equation}
In this situation, the variance takes the form 
\begin{eqnarray}  \label{dispdir}
\frac{1}{4} \mathcal{D} \approx \sum_{i=1}^Z \langle Z,Z| \frac{1}{ r_i^2}%
|Z,Z\rangle - \sum_{i=1}^Z \langle Z,Z| \frac{1}{r_i}|Z,Z\rangle^2.
\end{eqnarray}

To calculate the exchange contribution, it is necessary to know the
two-particle density of the electron distribution in the atom.
The exchange effects are taken into account in the Hartree-Fock method and
its relativistic versions by averaging the operator over the antisymmetrized
electron wave function. The Slater determinant, 
\begin{equation}
\Psi _{\alpha _{1}\alpha _{2}\ldots \alpha _{N}}^{\sigma _{1}\sigma
_{2}\ldots \sigma _{N}}(\mathbf{x}_{1},\mathbf{x}_{2},\ldots ,\mathbf{x}%
_{N})=\frac{1}{\sqrt{N!}}\epsilon ^{s_{1}s_{2}...s_{N}}\psi _{\alpha
_{s_{1}}}^{\sigma _{1}}(\mathbf{x}_{1})\psi _{\alpha _{s_{2}}}^{\sigma _{2}}(%
\mathbf{x}_{2})\ldots \psi _{\alpha _{s_{N}}}^{\sigma _{N}}(\mathbf{x}_{N}),
\label{slater}
\end{equation}%
represents the shell wave function, where $\psi _{\alpha _{s}}^{\sigma }(%
\mathbf{x)}$ are the wave functions of electrons; index $i=1,...,N$ numbers
the spatial coordinates and bispinor indices of the orbitals; and index $%
\alpha $ represents the quantum numbers of orbitals. In the case under
study, $\alpha =(njlm)$, where $n$ is the principal~quantum number, $j$ is
the total angular momentum, $l=j\pm 1/2\geq 0$ is the orbital angular
momentum, and $m=-j,\ldots ,j$ is the projection of the total angular
momentum on the $z$ axis. The Slater determinant is defined as a convolution
of the Levi-Civita symbol $\epsilon ^{s_{1}s_{2}\ldots s_{N}}=\pm 1$ with
the product of $N$ electron wave functions. The Levi-Civita symbol
antisymmetrizes the wave function by summing over $N!$ permutations. In the
one-determinant approximation, a fixed set of quantum numbers $(\alpha
_{1}\alpha _{2}\ldots \alpha _{N})$ completely describes the state of the
shell. The orthonormality condition for the wave functions is expressed as 
\begin{equation*}
\int d\mathbf{x}\psi _{\alpha }^{\dag }(\mathbf{x})\psi _{\beta }(\mathbf{x}%
)=\delta _{\alpha \beta }.
\end{equation*}

The calculations use a two-particle density matrix $i\neq k$, which is found
by integrating the product $\Psi _{\alpha _{1}\alpha _{2}\ldots \alpha
_{N}}^{\sigma _{1}\sigma _{2}\ldots \sigma _{N}}(\mathbf{x}_{1},\mathbf{x}%
_{2},$ $\ldots ,\mathbf{x}_{N})\Psi _{\alpha _{1}\alpha _{2}\ldots \alpha
_{N}}^{\bar{\sigma}_{1}\bar{\sigma _{2}}\ldots \bar{\sigma}_{N}\ast }(\bar{%
\mathbf{x}}_{1},\bar{\mathbf{x}}_{2},\ldots ,\bar{\mathbf{x}}_{N})$ over the
coordinates for $ \mathbf{x}_{h} = \bar{\mathbf{x}}_{h}$ and
contracting the bispinor indices $\sigma _{h}$ and $\bar{\sigma}_{h}$ of
orbitals with numbers $h$ other than $i$ and $k$: 
\begin{equation}
\rho _{\alpha _{1}\alpha _{2}\ldots \alpha _{N}}^{\sigma _{i}\tau _{k}\bar{%
\tau}_{k}\bar{\sigma}_{i}}(\mathbf{x}_{i},\mathbf{x}_{k};\mathbf{\bar{x}}%
_{k},\mathbf{\bar{x}}_{i})=\frac{1}{N(N-1)}\left( \delta ^{rr^{\prime
}}\delta ^{ss^{\prime }}-\delta ^{rs^{\prime }}\delta ^{sr^{\prime }}\right)
\psi _{\alpha _{r}}^{\sigma _{i}}(\mathbf{x}_{i})\psi _{\alpha _{s}}^{\tau
_{k}}(\mathbf{x}_{k})\psi _{\alpha _{s^{\prime }}}^{\bar{\tau}_{k}\ast }(%
\mathbf{\bar{x}}_{k})\psi _{\alpha _{r^{\prime }}}^{\bar{\sigma}_{i}\ast }(%
\mathbf{\bar{x}}_{i}).  \label{rho2}
\end{equation}%
The indices $r,s,r^{\prime },s^{\prime }=1,\ldots ,N$ are summed. Assuming $%
\mathbf{y}=\mathbf{\bar{y}}$, integrating over the variable $\mathbf{y}$ and
further contracting the spinor indices $\tau $ and $\bar{\tau}$, we find a
single-particle density matrix: 
\begin{equation}
\rho _{\alpha _{1}\alpha _{2}\ldots \alpha _{N}}^{\sigma \bar{\sigma}}(%
\mathbf{x};\mathbf{\bar{x}})=\frac{1}{N}\delta ^{rr^{\prime }}\psi _{\alpha
_{r}}^{\sigma }(\mathbf{x})\psi _{\alpha _{r^{\prime }}}^{\bar{\sigma}\ast }(%
\mathbf{\bar{x}}),  \label{rho1}
\end{equation}%
which is used to calculate the average values of single-particle operators.

Let's represent the upper and lower components of the electron wave functions as 
products of the radial and angular parts: 
\begin{equation}
\psi _{\alpha }(\mathbf{x})=\left( 
\begin{array}{l}
f_{_{\varkappa }}(r)\Omega _{j_{\alpha }m_{\alpha }}^{l_{\alpha }}(\mathbf{n}%
) \\ 
ig_{_{\varkappa }}(r)\Omega _{j_{\alpha }m_{\alpha }}^{l_{\alpha }^{\prime }}(%
\mathbf{n})%
\end{array}%
\right) .  \label{deco}
\end{equation}%
Here $f_{_{\varkappa }}(r)$ and $g_{_{\varkappa }}(r)$ are real functions, 
$\alpha =(njlm),$ $\varkappa =(njl)$, $l^{\prime} = 2j - l$, $\mathbf{n}=\mathbf{x}/|\mathbf{x}|$,
and $\Omega _{jm}^{l}(\mathbf{n})$ is the spherical spinor, describing a spin--%
$1/2$ particle. Spherical spinors are normalized by the condition 
\cite{Varshalovich:1988} 
\begin{equation}
\int do_{\mathbf{n}}\Omega _{jm}^{l}(\mathbf{n})^{\dagger }\Omega
_{j^{\prime }m^{\prime }}^{l^{\prime }}(\mathbf{n})=\delta ^{ll^{\prime
}}\delta _{jj^{\prime }}\delta _{mm^{\prime }}.  \notag
\end{equation}


\begin{table}[t]
\scriptsize
\centering
\caption{
Matrix elements $\langle \varkappa |r^{-1}|\bar{\varkappa}\rangle$ and $\langle \varkappa |r^{-2}|\bar{\varkappa}\rangle$ 
with $\varkappa = (njl)$ and $\bar{\varkappa} = (\bar{n}jl)$ for orbital states of electrons in calcium and germanium atoms. Numerical values are given in atomic units. The computations use non-relativistic Roothaan-Hartree-Fock wave functions \cite{Clementi:1974}.  The final two lines depict the diagonal matrix elements of $\varkappa = \bar{\varkappa}$ calculated by the relativistic Dirac-Fock method \cite{Desclaux:1973}. The higher and lower lines represent the total angular momentum of $j=l\pm1/2$, respectively.
}
\vspace{2mm}
\label{tab:table20}
\begin{tabular}{|c|}
\hline
\hline
$_{20}$Ca \\
\begin{tabular}{c|rrrr|r|rrr|r|r}
\hline
$\langle \varkappa |r^{-1}|\bar{\varkappa}\rangle$ & 1S~~      & 2S~~      & 3S~~     & 4S~~    &    & 2P   & 3P~~     & 4P~~     &    & 3D~~   \\
\hline
1S & 19.54   & 3.458   & 1.158  & 0.275~ &        &         &        &         &       &          \\ 
2S &         & 4.041   & 0.776 & 0.179~  &   ~2P~ & 3.944   & -0.634 &  $\;\;\;\;\;\;$~       &       &     \\
3S &         &         & 1.189  & 0.150~ &   ~3P~ &         & 1.059  &         &       &       \\
4S &         &         &        & 0.300~ &        &         &        &         &       &         \\
\hline
\cite{Desclaux:1973}& 19.75  & 4.089   & 1.198   &  0.301~ &    & 3.947 & 1.059 &   &    &         \\
&         &         &        &         &                        & 3.987 & 1.066  &   &    &         \\
\hline
$\langle \varkappa |r^{-2}|\bar{\varkappa}\rangle$ & 1S~~      & 2S~~      & 3S~~     & 4S~~     &    & 2P~~      & 3P~~     & 4P~~   &    & 3D~~  \\ 
\hline
1S & 769.4   & 200.4   & 68.52   & 16.35~ &      &         &        &           &    &                          \\ 
2S &         & 67.04   & 21.80   & 5.185~ & ~2P~ & 21.49   & -5.373 &   &    &                           \\ 
3S &         &         & 8.395   & 1.921~ & ~3P~ &         &  2.462 &     & $\;\;\;\;\;\;$ &  $\;\;\;\;\;\;$~          \\ 
4S &         &         &         & 0.535~ &      &         &        &     &     &                  \\ 
\hline
\cite{Desclaux:1973}& 794.3  & 70.01   & 8.752 &  0.557~      &    & 21.54 & 2.447  &  &    &              \\
&         &         &        &         &                          & 22.24 & 2.541 &  &    &             \\
\hline
\hline
\multicolumn{11}{c}{$_{32}$Ge} \\
\hline
$\langle \varkappa |r^{-1}|\bar{\varkappa}\rangle$ 
   & 1S~~      & 2S~~      & 3S~~     & 4S~~      &    & 2P~~      & 3P~~     & 4P~~      &    & 3D~~     \\
\hline
1S & 31.51   & 5.917   & -2.203 & -0.564~    &      &         &        &         &     &                    \\
2S &         &  6.950  & -1.477 & -0.366~    & ~2P~ & 6.890   & 1.248  & 0.231~  &     &                         \\
3S &         &         & 2.233  & 0.310~     & ~3P~ &         & 2.104  & 0.223~  & ~3D~ & 1.793~        \\
4S &         &         &        &  0.580~    & ~4P~ &         &        & 0.437~ &      &                 \\
\hline
\cite{Desclaux:1973}& 32.28 & 7.179 & 2.289 & 0.591~ &    & 6.917 & 2.112 &  0.441~ & & 1.783~   \\
                    &       &       &       &        &    & 7.107 & 2.156 &  0.448~ & & 1.796~    \\
\hline
$\langle \varkappa |r^{-2}|\bar{\varkappa}\rangle$ & 1S~~      & 2S~~      & 3S~~     & 4S~~     &    & 2P~~      & 3P~~     & 4P~~   &   & 3D~~  \\ 
\hline
1S & 1995.   & 555.1   & -211.8  & -54.40~           &      &         &        &           &    &                           \\ 
2S &         &  197.2  & -71.55  & -18.31~           & ~2P~ & 64.75   & 18.57  & 3.527~    &    &                           \\ 
3S &         &         &  30.39  & 7.508~            & ~3P~ &         & 9.613  & 1.658~    & ~3D~ & 4.278~            \\ 
4S &         &         &         & 2.206~            & ~4P~ &         &        & 0.497~    &     &                  \\ 
\hline
\cite{Desclaux:1973}& 2168. & 221.3 & 34.18 & 2.482~ &      & 65.39 & 9.749 & 0.507~ &    & 4.233~             \\
&         &         &        &         &                    & 71.41 & 10.61 & 0.547~ &    & 4.301~             
\end{tabular} \\
\hline
\hline
\end{tabular}
\end{table}

The average value of the sum of the products of single-particle operators $%
\mathcal{O}^{\bar{\sigma}\sigma }(r_{i})\mathcal{O}^{\bar{\tau}\tau }(r_{k})$
is given by the expression 
\begin{equation}
\sum_{i\neq k}\langle Z,N|\hat{\mathcal{O}}(r_{i})\hat{\mathcal{O}}%
(r_{k})|Z,N\rangle =\sum_{i\neq k}\int d\mathbf{x}_{i}d\mathbf{x}_{k}%
\mathcal{O}^{\bar{\sigma}\sigma }(r_{i})\mathcal{O}^{\bar{\tau}\tau
}(r_{k})\rho ^{\sigma \tau \bar{\tau}\bar{\sigma}}(\mathbf{x}_{i},\mathbf{x}%
_{k};\mathbf{x}_{k},\mathbf{x}_{i}).
\end{equation}%
In the case we are interested in, $\mathcal{O}^{\bar{\sigma}\sigma
}(r)=\delta ^{\bar{\sigma}\sigma }\mathcal{O}(r)$. Substituting the
expression (\ref{deco}) into the right side of the equation, followed by
integration over the angular variables and taking into account the orthonormality
of spherical spinors gives 
\begin{eqnarray}
\sum_{i\neq k}\langle Z,N|\hat{\mathcal{O}}(r_{i})\hat{\mathcal{O}}%
(r_{k})|Z,N\rangle  &=&\sum_{i\neq k}\int r_{i}^{2}dr_{i}r_{k}^{2}dr_{k}%
\mathcal{O}(r_{i})\mathcal{O}(r_{k})\sum_{\alpha _{i}\alpha _{k}}\frac{1}{%
N(N-1)}  \label{lesslong} \\
&\times &\left( \mathcal{P}_{\varkappa _{i}\varkappa _{i}}(r_{i})\mathcal{P}%
_{\varkappa _{k}\varkappa _{k}}(r_{k})-\delta ^{l_{i}l_{k}}\delta
_{j_{i}j_{k}}\delta _{m_{i}m_{k}}\mathcal{P}_{\varkappa _{i}\varkappa
_{k}}(r_{i})\mathcal{P}_{\varkappa _{k}\varkappa _{k}}(r_{k})\right) , 
\notag
\end{eqnarray}%
where 
\begin{equation*}
\mathcal{P}_{\varkappa _{i}\varkappa _{k}}(r)=f_{\varkappa
_{i}}(r)f_{\varkappa _{k}}(r)+g_{\varkappa _{k}}(r)g_{\varkappa _{i}}(r).
\end{equation*}%
The summation by electron states includes the terms $\varkappa
_{i}=\varkappa _{k}$, which correspond to the common shell of electrons $i$
and $k$. As a consequence of the antisymmetry of the total wave function,
the direct and exchange contributions at $m_{i}=m_{k}$ cancel each other.
The diagonal terms $\alpha _{i}=\alpha _{k}$ could be excluded from the sum,
however, calculations are simplified when the summation over the quantum
numbers of two electrons is uncorrelated, i.e., it includes the diagonal.


\begin{table}[]
\scriptsize
\centering
\caption{
Matrix elements $\langle \varkappa |r^{-1}|\bar{\varkappa}\rangle$ and $\langle \varkappa |r^{-2}|\bar{\varkappa}\rangle$ for the orbital states of electrons in selenium and zirconium atoms. Notations are identical to those in Table ~\ref{tab:table20}.
}
\vspace{2mm}
\label{tab:table34}
\begin{tabular}{|c|}
\hline
\hline
$_{34}$Se \\
\begin{tabular}{c|rrrrr|r|rrr|r|rr}
\hline
$\langle \varkappa |r^{-1}|\bar{\varkappa}\rangle$ & 1S~~      & 2S~~      & 3S~~     & 4S~~      & 5S~~    &    & 2P~~   & 3P~~     & 4P~~     &    & 3D~~          & 4D~~ \\
\hline
1S & 33.50   & 6.325   & -2.403 & -0.699 &   $\;\;\;\;\;\;\;\;$~      &      &         &        &         &    &           &    $\;\;\;\;\;\;\;\;$~  \\ 
2S &         & 7.434   & -1.610 & -0.452 &         & ~2P~ & 7.379   & 1.368  & 0.313~  &     &             &              \\
3S &         &         & 2.433  &  0.379 &         & ~3P~ &         & 2.311  & 0.298~  & ~3D~ & 2.040     &     \\
4S &         &         &        &  0.695 &         & ~4P~ &         &        & 0.563~  &   &           &        \\
\hline
\cite{Desclaux:1973}& 34.57  & 7.714   & 2.502   & 0.711  &         &    & 7.412 & 2.322 & 0.561~ &    & 2.029   &       \\
&         &         &        &         &         &                       & 7.645 & 2.375 & 0.573~ &    & 2.044  &      \\
\hline
$\langle \varkappa |r^{-2}|\bar{\varkappa}\rangle$ & 1S~~      & 2S~~      & 3S~~     & 4S~~      & 5S~~      &    & 2P~~      & 3P~~     & 4P~~      &    & 3D~~          & 4D~~          \\ 
\hline
1S & 2255.   & 631.2   & -245.8  & -71.72  &         &      &         &        &           &    &             &              \\ 
2S &         & 225.4   & -83.48  & -24.26  &         & ~2P~ & 74.18   & 21.83  & 5.137~    &    &             &              \\ 
3S &         &         & 36.13   &  10.13  &         & ~3P~ &         & 11.55  & 2.459~    & ~3D~ & 5.443     &       \\ 
4S &         &         &         &  3.329  &         & ~4P~ &         &        & 0.857~    &     &           &        \\ 
\hline
\cite{Desclaux:1973}& 2479.  & 257.1   & 41.26 &   3.809 &     &    & 75.05 & 11.73 & 0.861~ &    & 5.395       &       \\
&         &         &        &         &         &                  & 82.97 & 12.92 & 0.944~ &    & 5.483      &       \\
\hline
\hline
\multicolumn{13}{c}{$_{40}$Zr} \\
\hline
$\langle \varkappa |r^{-1}|\bar{\varkappa}\rangle$ 
& 1S~~      & 2S~~      & 3S~~     & 4S~~      & 5S~~      &    & 2P~~      & 3P~~     & 4P~~      &    & 3D~~          & 4D~~ \\
\hline
1S & 39.49   & -7.553  & -3.023 &  1.134 &  0.301~ &      &         &        &         &     &             &              \\
2S &         &  8.891  &  2.022 & -0.728 & -0.192~ & ~2P~ & 8.849   & -1.735 & -0.563~ &     &             &              \\
3S &         &         &  3.055 & -0.603 & -0.153~ & ~3P~ &         & 2.950  &  0.524~ & ~3D~ & 2.742     & -0.319~   \\
4S &         &         &        &  1.067 &  0.140~ & ~4P~ &         &        &  0.951~ &  ~4D~ &           & 0.628~      \\
5S &         &         &        &        &  0.306~ &      &         &        &         &     &             &              \\
\hline
\cite{Desclaux:1973}& 41.26 & 9.368 & 3.177  & 1.098  &   0.317~  &    & 8.909 & 2.972 &  0.954~ &    & 2.730     & 0.606~       \\
&         &         &        &         &         &                     & 9.307 & 3.064 &  0.978~ &    & 2.755     & 0.612~       \\
\hline
$\langle \varkappa |r^{-2}|\bar{\varkappa}\rangle$ & 1S~~      & 2S~~      & 3S~~     & 4S~~      & 5S~~      &    & 2P~~      & 3P~~     & 4P~~      &    & 3D~~          & 4D~~           \\ 
\hline
1S & 3131.   & -889.0  & -365.2  &  137.7  &  36.48~ &      &         &        &           &    &             &              \\ 
2S &         &  321.8  &  125.7  & -47.15  & -12.49~ & ~2P~ & 106.4   & -33.32 & -11.01~   &    &             &              \\ 
3S &         &         & 57.09   & -20.66  & -5.458~ & ~3P~ &         &  18.61 &   5.617~  & ~3D~ & 9.569     & -1.720~      \\ 
4S &         &         &         &  8.537  &  2.178~ & ~4P~ &         &        &   2.562~  & ~4D~ &           &  0.745~      \\ 
5S &         &         &         &         & 0.657~  &      &         &        &           &    &             &              \\
\hline
\cite{Desclaux:1973}& 3577.  & 387.3   & 68.702 &   10.23 &  0.810~ &    & 108.2 & 19.06 & 2.616~ &    & 9.513      & 0.701~       \\
&         &         &        &         &         &                       & 124.8 & 21.83 & 2.972~ &    & 9.714      & 0.719~       
\end{tabular} \\
\hline
\hline
\end{tabular}
\end{table}

Let's assume each pair $(jl)$ has a maximum of one open shell for some $n
$. When summing over projections $m_{i}$ and $m_{k}$, one should keep in
mind that if $m_{i}$ takes the values from the interval $-j,\ldots ,j$, $%
m_{k}$ can only take some of them. The second line of Eq.~(\ref{lesslong})
yields a multiplier of $(2j_{i}+1)\mathcal{N}_{\varkappa _{k}}$, where $%
\mathcal{N}_{\varkappa }\leq 2j+1$ represents the number of electrons on a
completely or partially filled shell $\varkappa $. For completely filled
shells, $\mathcal{N}_{\varkappa }=2j+1$. In the third line, summation by $%
m_{i}$ removes the Kronecker symbol $\delta _{m_{i}m_{k}}$. Summation by $%
m_{k}$ yields the multiplier $\mathcal{N}_{\varkappa _{k}}$. The expression
below is obtained by summing over projections $m_{i}$ and $m_{k}$,
integrating radial variables $r_{i}$ and $r_{k}$, and summing over $i\neq k$%
: 
\begin{equation}
\sum_{i\neq k}\langle Z,N|\hat{\mathcal{O}}(r_{i})\hat{\mathcal{O}}%
(r_{k})|Z,N\rangle =(\sum_{\varkappa }\mathcal{N}_{\varkappa }\langle
\varkappa |\mathcal{O}(r)|\varkappa \rangle )^{2}-\sum_{\varkappa \bar{%
\varkappa}}\min (\mathcal{N}_{{\varkappa }},\mathcal{N}_{\bar{\varkappa}%
})\langle {\varkappa }|\mathcal{O}(r)|\bar{\varkappa}\rangle ^{2},
\label{fin}
\end{equation}%
where $\varkappa =(njl)$, $\bar{\varkappa}=(\bar{n}jl)$ and 
\begin{equation}
\langle \varkappa |\mathcal{O}(r)|\bar{\varkappa}\rangle =\int r^{2}dr%
\mathcal{O}(r)\mathcal{P}_{\varkappa \bar{\varkappa}}(r).  \label{lesslong1}
\end{equation}%
This equation is appropriate for calculating the average values of operators
that do not mix the upper and lower components of bispinors. 
Summing the same terms in Eq.~(\ref{lesslong}) cancels the multiplier $1/(N(N-1))$.
In the second term on the right side of Eq.~(\ref{fin}), the
summation is carried out over the quantum numbers $n,\bar{n},j$, and $l$.
The averaging is performed over the radial density of
electrons in the parent atom $N=Z$. In non-relativistic theory where $g_{\varkappa}(r) = 0$, the radial
part of wave functions $f_{\varkappa}(r)$ is exclusively dependent on $n$ and $l$. For completely
filled shells, summation by $j=l\pm 1/2$ enables the replacement of $%
|njl\rangle \rightarrow |nl\rangle $ and $(2j+1)\rightarrow 2(2l+1)$.

A single-particle density matrix is used to calculate the average value of
the sum of single-particle operators $\hat{\mathcal{O}}(r_{i})$: 
\begin{equation}  \label{mean}
\sum_{i}\langle Z,N| \hat{\mathcal{O}}(r_{i})|Z,N\rangle =\sum_{\varkappa}%
\mathcal{N}_{\varkappa}\langle \varkappa|\mathcal{O}(r)| \varkappa \rangle .
\end{equation}

The variance consists of the sum of the diagonal and off-diagonal components 
$\langle Z,N|\hat{\mathcal{O}}(r_{i})\hat{\mathcal{O}}(r_{k})|Z,N\rangle $
minus the square of the sum of the components $\langle Z,N|\hat{\mathcal{O}}%
(r_{i})|Z,N\rangle $. In the case under consideration, $\mathcal{O}(r)=r^{-1}
$. The first and third terms are calculated using Eq.~(\ref{mean}), the
second term is given by Eq.~(\ref{fin}). We remark that the third term is
cancelled by the first one on the right side of Eq.~(\ref{fin}). Finally,
the variance of the excitation energy of the electron shell turns out to
be
\begin{equation}
\frac{1}{4}\mathcal{D}=\sum_{\varkappa }\mathcal{N}_{\varkappa }\langle
\varkappa |{r^{-2}}|\varkappa \rangle -\sum_{\varkappa \bar{\varkappa}}\min (%
\mathcal{N}_{{\varkappa }},\mathcal{N}_{\bar{\varkappa}})\langle {\varkappa }%
|{r}^{-1}|\bar{\varkappa}\rangle ^{2}.  \label{D4F}
\end{equation}%
The sum of the diagonal components coincides with the right side of Eq.~(\ref%
{dispdir}). The off-diagonal components in the second term are associated
with exchange effects. The exchange effects reduce the variance.


\begin{table}[]
\scriptsize
\centering
\caption{
Matrix elements $\langle \varkappa |r^{-1}|\bar{\varkappa}\rangle$ and $\langle \varkappa |r^{-2}|\bar{\varkappa}\rangle$ for the orbital states of electrons in molybdenum and cadmium atoms. Notations are identical to those in Table ~\ref{tab:table20}.
}
\vspace{2mm}
\label{tab:table40}
\begin{tabular}{|c|}
\hline
\hline
$_{42}$Mo \\
\begin{tabular}{c|rrrrr|r|rrr|r|rr}
\hline
$\langle \varkappa |r^{-1}|\bar{\varkappa}\rangle$ & 1S~~      & 2S~~      & 3S~~     & 4S~~      & 5S~~     &    & 2P~~      & 3P~~     & 4P~~     &    & 3D~~          & 4D~~ \\
\hline
1S & 41.49   & 7.962   &  3.231 & -1.255 & 0.321~  &      &         &        &           &      &             &              \\
2S &         & 9.378   &  2.160 & -0.803 & 0.204~  & ~2P~ &  9.339  & -1.858 & -0.626~   &      &             &              \\
3S &         &         &  3.264 & -0.665 & 0.163~  & ~3P~ &         &  3.164 &  0.582~   & ~3D~ & 2.970       & -0.361~       \\
4S &         &         &        &  1.171 &-0.149~  & ~4P~ &         &        &  1.052~   & ~4D~ &             &  0.714~       \\
5S &         &         &        &        & 0.327   &      &         &        &           &      &             &              \\
\hline
\cite{Desclaux:1973}& 43.55 &  9.939 & 3.409  & 1.209  &   0.322~  &    & 9.412 & 3.190 &  1.059~ &    & 2.958     & 0.695~       \\
&         &         &        &         &         &                      & 9.879 & 3.300 &  1.089~ &    & 2.987     & 0.705~       \\
\hline
$\langle \varkappa |r^{-2}|\bar{\varkappa}\rangle$ & 1S~~      & 2S~~      & 3S~~     & 4S~~      & 5S~~      &    & 2P~~      & 3P~~     & 4P~~      &    & 3D~~          & 4D~~          \\ 
\hline
1S & 3455.   &  984.9  & 410.3  & -160.1   & 40.94~   &      &         &        &          &    &             &              \\ 
2S &         &  357.8  & 141.7  & - 55.02  & 14.06~   & ~2P~ & 118.4   & -37.69 & -13.17~   &    &             &              \\ 
3S &         &         &  65.20 & - 24.42  &  6.223~  & ~3P~ &         &  21.34 &   6.697~  & ~3D~ & 11.17      & -2.120~       \\ 
4S &         &         &         &   10.41 & -2.564~  & ~4P~ &         &        &   3.157~  & ~4D~ &             & 0.965~       \\ 
5S &         &         &         &         &  0.748~  &      &         &        &          &    &             &              \\
\hline
\cite{Desclaux:1973}& 4005. & 439.4 & 80.03  &   12.73 &  0.830~ &    & 120.7  & 21.93 & 3.243~ &    & 11.11      & 0.930~       \\
&         &         &        &         &         &                     & 141.5  & 25.50 & 3.744~ &    & 11.37      & 0.960~    \\   
\hline
\hline
\multicolumn{13}{c}{$_{48}$Cd} \\
\hline
$\langle \varkappa |r^{-1}|\bar{\varkappa}\rangle$ & 1S~~      & 2S~~      & 3S~~     & 4S~~      & 5S~~    &    & 2P~~      & 3P~~     & 4P~~     &  & 3D~~         & 4D~~ \\
\hline
1S & 47.48   &   9.194 & -3.856 &-1.617 & 0.3840~  &      &         &        &          &      &           &              \\
2S &         &  10.84  & -2.575 &-1.030 & 0.2429~  & ~2P~ & 10.81   & -2.174 & -0.820~ &      &           &              \\
3S &         &         &  3.895 & 0.851 &-0.1933~  & ~3P~ &         &  3.774 &  0.756~  & ~3D~ & 3.637     & -0.532~      \\
4S &         &         &        & 1.485 &-0.1793~  & ~4P~ &         &        &  1.369~  & ~4D~ &           &  1.063~      \\
5S &         &         &        &       & 0.3905~  &      &         &        &          &      &           &              \\
\hline
\cite{Desclaux:1973}& 50.64  & 11.72   & 4.127   &  1.554 &  0.4128~ &    & 10.93 & 3.850 &  1.380~  &    & 3.630      & 1.049~       \\ 
&         &         &        &         &         &                        & 11.66 & 4.024 &  1.432~  &    & 3.676      & 1.066~       \\
\hline
$\langle \varkappa |r^{-2}|\bar{\varkappa}\rangle$ & 1S~~      & 2S~~      & 3S~~     & 4S~~      & 5S~~     &  & 2P~~  & 3P~~  & 4P~~      &    & 3D~~  & 4D~~  \\ 
\hline
1S & 4524.   & 1302.   & -561.4  & -236.5  & 56.21~   &    &         &        &          &      &             &              \\ 
2S &         & 477.6   & -195.6  & -81.99  & 19.47~   & ~2P~ & 158.4 & -51.48 & -20.11~  &      &             &              \\ 
3S &         &         &  92.91  &  37.55  & -8.893~  & ~3P~ &       &  29.99 &  10.44~  & ~3D~ & 16.62       & -3.886~      \\ 
4S &         &         &         &  17.16  & -3.939~  & ~4P~ &       &        &  5.374~  & ~4D~ &             &  2.098~      \\ 
5S &         &         &         &         &  1.071~  &      &       &        &          &      &             &              \\
\hline
\cite{Desclaux:1973} & 5504. & 672.8 & 121.9  &   22.50 &  1.490~ &    &  162.7 & 31.80  & 5.571~ &    & 16.55      & 2.058~       \\
&                 &        &         &         &                  &    &  201.4 & 38.96  & 6.775~ &    & 17.05      & 2.142~       
\end{tabular} \\
\hline
\hline
\end{tabular}
\end{table}


\begin{table}[]
\scriptsize
\centering
\caption{
Matrix elements $\langle \varkappa |r^{-1}|\bar{\varkappa}\rangle$ and $\langle \varkappa |r^{-2}|\bar{\varkappa}\rangle$ for the orbital states of electrons in tellurium and xenon atoms. Notations are identical to those in Table ~\ref{tab:table20}.
}
\vspace{2mm}
\label{tab:table52}
\begin{tabular}{|c|}
\hline
\hline
$_{52}$Te\\
\begin{tabular}{c|rrrrr|r|rrrr|r|rr}
\hline
$\langle \varkappa |r^{-1}|\bar{\varkappa}\rangle$ & 1S~~      & 2S~~      & 3S~~     & 4S~~      & 5S~~    &    & 2P~~      & 3P~~     & 4P~~      & 5P~~     &    & 3D~~          & 4D~~          \\ 
\hline
1S & 51.47   & -10.02 & -4.273  &  1.889 &  0.639~ &      &         &        &         &          &      &             &              \\ 
2S &         &  11.82 &  2.851  & -1.201 & -0.403~ & ~2P~ & 11.80   & -2.463 & -0.971 &  0.287~ &      &             &              \\ 
3S &         &         & 4.316  & -0.991 & -0.318~ & ~3P~ &         &  4.230 &  0.902 & -0.256~ & ~3D~ & 4.084       & 0.684~       \\ 
4S &         &         &        &  1.722 &  0.290~ & ~4P~ &         &        &  1.616 & -0.244~ & ~4D~ &             & 1.369~       \\ 
5S &         &         &        &        &  0.590~ & ~5P~ &         &        &         &  0.504~ &      &             &              \\
\hline
\cite{Desclaux:1973}& 55.57 & 12.96 & 4.627 &  1.815 &  0.602~ &     & 11.95  & 4.294 &  1.631 &  0.470~ &    & 4.073      & 1.354~        \\
                    &         &         &        &         &         &    & 12.91  & 4.526 &  1.700 &  0.493~ &    & 4.133      & 1.374~       \\
\hline
$\langle \varkappa |r^{-2}|\bar{\varkappa}\rangle$ & 1S~~      & 2S~~      & 3S~~     & 4S~~      & 5S~~      &    & 2P~~      & 3P~~     & 4P~~      & 5P~~      &    & 3D~~  & 4D~~          \\ 
\hline
1S & 5316.   & -1539.  & -674.9  &  300.0  &  101.6~ &      &         &        &         &         &      &             &              \\
2S &         &  567.1  &  236.4  & -104.5  & -35.35~ & ~2P~ & 188.4   & -63.40 & -26.07  &  7.759~ &      &             &              \\
3S &         &         &  114.1  &  -48.63 & -16.40~ & ~3P~ &         & 37.75  & 13.91   & -4.107~ & ~3D~ & 20.80       & 5.625~        \\
4S &         &         &         &   23.35 &  7.628~ & ~4P~ &         &        &  7.485  & -2.026~ & ~4D~ &             & 3.401~        \\
5S &         &         &         &         &  2.841~ & ~5P~ &         &        &         &  0.806~ &       &             &              \\
\hline
\cite{Desclaux:1973}& 6713. & 785.3 & 157.7 &   32.09 & 3.731~ &      & 194.6  & 39.51   & 7.809  &  0.719~ &    & 20.73      & 3.347~       \\
 &         &         &        &         &         &                   & 251.2  & 50.41   &  9.852 &  0.916~ &    & 21.48      & 3.478~       \\
\hline
\hline
\multicolumn{14}{c}{$_{54}$Xe}  \\
\hline
$\langle \varkappa |r^{-1}|\bar{\varkappa}\rangle$ & 1S~~      & 2S~~      & 3S~~     & 4S~~      & 5S~~     &    & 2P~~      & 3P~~     & 4P~~      & 5P~~     &    & 3D~~          & 4D~~          \\ 
\hline
1S & 53.47   & 10.43   & -4.482 &  2.027   &  0.712~  &      &         &        &         &         &    &             &              \\ 
2S &         & 12.31   & -2.990 &  1.287   &  0.448~  & ~2P~ & 12.29   & 2.584  & -1.047  & -0.317~ &    &             &              \\ 
3S &         &         &  4.527 & -1.063   & -0.353~  & ~3P~ &         & 4.445  & -0.972  & -0.282~ & ~3D~ & 4.304   & 0.751~      \\ 
4S &         &         &        &  1.843   &  0.321~  & ~4P~ &         &        &  1.741  &  0.267~ & ~4D~ &             & 1.509~       \\ 
5S &         &         &        &          &  0.648~  & ~5P~ &         &        &         &  0.547~ &    &             &              \\
\hline
\cite{Desclaux:1973}& 58.11 & 13.61 & 4.884 &  1.950 & 0.681~ &        & 12.47 & 4.519 &  1.760 &  0.546~ &    & 4.293      & 1.495~       \\
   &         &         &        &         &         &                   & 13.55 & 4.784 &  1.841 &  0.576~ &    & 4.362      & 1.517~       \\
\hline
$\langle \varkappa |r^{-2}|\bar{\varkappa}\rangle$ & 1S~~      & 2S~~      & 3S~~     & 4S~~      & 5S~~     &    & 2P~~      & 3P~~     & 4P~~      & 5P~~    &    & 3D~~          & 4D~~          \\ 
\hline
1S & 5735.   & 1665.   & -735.7 &  334.7 &  117.6~  &      &         &        &         &         &    &             &              \\ 
2S &         & 614.8   & -258.2 &  116.8 &  41.01~  & ~2P~ & 204.4   & 69.35  & -29.34  & -8.935~ &    &             &              \\ 
3S &         &         &  125.5 & -54.76 & -19.16~  & ~3P~ &         & 41.60  & -15.78  & -4.761~ & ~3D~ & 23.06     & 6.526~      \\ 
4S &         &         &        &  26.85 &   9.107~ & ~4P~ &         &        &   8.687 &  2.405~ & ~4D~ &             & 4.097~       \\ 
5S &         &         &        &        &   3.506~ & ~5P~ &         &        &         &  0.970~ &    &             &              \\
\hline
\cite{Desclaux:1973}& 7391. & 879.8 & 178.5 &   37.89 &  5.020~ &    & 211.7 & 43.73 &  9.104 &  0.998~ &    & 22.99      & 4.048~       \\
   &         &         &        &         &         &                   & 279.6 & 37.05 & 11.72  &  1.305~ &    & 23.89      & 4.212~       \\
\end{tabular}\\
\hline
\hline
\end{tabular}
\end{table}


\begin{table}[]
\scriptsize
\centering
\caption{
Matrix elements $\langle njl |r^{-1}|\bar{n} jl\rangle$ and $\langle njl|r^{-2}|\bar{n} jl\rangle$ for orbital states of electrons in neodymium and uranium atoms, obtained by the DHF method with the use of the program package G\textsc{rasp}2018  \cite{Fischer:2019}. Each matrix element except for $l=0$ is given 
two values of the total angular momentum $j$. The upper and lower lines correspond to $j=l\pm1/2$.
}
\vspace{2mm}
\label{tab:table60}
\begin{tabular}{|c|}
\hline
\hline
$_{60}$Nd\\
\begin{tabular}{c|rrrrrrr|r|rrrrr|r|rrrr|r|rr}
\hline
$\langle \varkappa |r^{-1}|\bar{\varkappa}\rangle$ & 1S~~  & 2S~~  & 3S~~  & 4S~~ & 5S~~ & 6S~~ & 7S~~ &  &  2P~~  & 3P~~ & 4P~~ & 5P~~ & 6P~~ & ~~ & 3D~~ & 4D~~ & 5D~~ & 6D~~ & & 4F~~ & 5F~~      \\ 
\hline
1S & 66.04 & 14.07 & 6.167 & 2.894 & 1.132 & 0.299 &  & 2P & 14.02 & 3.055 & 1.289 & 0.443 &  & 3D & ~ 4.959 & 0.914 &  &  & 4F &~1.291 &      \\ 
2S &       & 15.65 & 4.074 & 1.825 & 0.707 & 0.186 &  &    & 15.59 & 3.524 & 1.497 & 0.529 &   &    & 5.055 & 0.943 &  &  &    &~1.306 &     \\ 
3S &       &       & 5.689 & 1.439 & 0.533 & 0.139 &  & 3P &       & 5.202 & 1.198 & 0.393 &   & 4D &       & 1.845 &  &  & 5F &  &      \\ 
4S &       &       &       & 2.336 & 0.464 & 0.117 &  &    &       & 5.590 & 1.322 & 0.447 &   &    &       & 1.877 &  &  &    &  &     \\ 
5S &       &       &       &       & 0.874 & 0.112 &  & 4P &       &       & 2.105 & 0.368 &   & 5D &       &       &  &  &    &  &      \\
6S &       &       &       &       &       & 0.258 &  &    &       &       & 2.227 & 0.405 &   &    &       &       &  &  &    &  &      \\
7S &       &       &       &       &       &       &  & 5P &       &       &       & 0.725 &   & 6D &       &       &  &  &    &  &      \\
   &       &       &       &       &       &       &  &    &       &       &       & 0.773 &   &    &       &       &  &  &    &  &      \\
   &       &       &       &       &       &       &  & 6P &       &       &       &       &   &    &       &       &  &  &    &  &      \\
   &       &       &       &       &       &       &  &    &       &       &       &       &   &    &       &       &  &  &    &  &       \\
   
\hline
$\langle \varkappa |r^{-2}|\bar{\varkappa}\rangle$ & 1S~~ & 2S~~ & 3S~~  & 4S~~ & 5S~~ & 6S~~ & 7S~~ &  &  2P~~  & 3P~~ & 4P~~ & 5P~~ & 6P~~ &  & 3D~~ & 4D~~ & 5D~~ & 6D~~ & & 4F~~ & 5F~~      \\ 
\hline
1S & 9769. & 3121. & 1413. & 667.8 & 261.6 & 69.13 &  & 2P & 268.0 & 94.64 & 41.87 & 14.51 &   & 3D & 30.55 & 9.217 &  &  & 4F &~2.209 &      \\ 
2S &       & 1204. & 525.9 & 247.6 & 96.90 & 25.61 &  &    & 382.4 & 142.9 & 63.93 & 22.82 &   &    & 32.02 & 9.784 &  &  &    &~2.255 &     \\ 
3S &       &       & 255.0 & 117.0 & 45.69 & 12.07 &  & 3P &       & 57.93 & 23.12 & 7.934 &   & 4D &       & 6.097 &  &  & 5F &  &      \\ 
4S &       &       &       & 58.22 & 22.24 & 5.868 &  &    &       & 81.42 & 33.62 & 11.91 &   &    &       & 6.410 &    &  &  &  &     \\ 
5S &       &       &       &       & 9.219 & 2.377 &  & 4P &       &       & 13.08 & 4.139 &   & 5D &       &       &    &  &  &  &      \\
6S &       &       &       &       &       & 0.686 &  &    &       &       & 18.13 & 6.028 &   &    &       &       &    &  &  &  &      \\
7S &       &       &       &       &       &       &  & 5P &       &       &       & 1.821 &   & 6D &       &       &    &  &  &  &      \\
   &       &       &       &       &       &       &  &    &       &       &       & 2.579 &   &    &       &       &    &  &  &  &      \\
   &       &       &       &       &       &       &  & 6P &       &       &       &       &   &    &       &       &    &  &  &  &      \\
   &       &       &       &       &       &       &  &    &       &       &       &       &   &    &       &       &    &  &  &  &       \\

\hline
\hline
\multicolumn{22}{c}{$_{92}$U}  \\
\hline
$\langle \varkappa |r^{-1}|\bar{\varkappa}\rangle$ & 1S~~  & 2S~~  & 3S~~  & 4S~~ & 5S~~ & 6S~~ & 7S~~ &  &  2P~~  & 3P~~ & 4P~~ & 5P~~ & 6P~~ &  & 3D~~ & 4D~~ & 5D~~ & 6D~~ & & 4F~~ & 5F~~      \\ 
\hline
1S & 122.4 & 30.38 & 14.02 & 7.179 & 3.541 & 1.565 & 0.464 ~& 2P &~22.78 & 5.335 & 2.478 & 1.150 & 0.445 ~& 3D & 8.493 & 1.778 & 0.686 & 0.189 ~& 4F &~3.326 & 0.419     \\ 
2S &       & 30.96 & 9.205 & 4.519 & 2.207 & 0.973 & 0.288~ &    & 30.88 & 8.048 & 3.801 & 1.785 & 0.734 ~&    & 8.924 & 1.909 & 0.741 & 0.204 ~&    &~3.384 & 0.428    \\ 
3S &       &       & 11.28 & 3.348 & 1.562 & 0.682 & 0.202 ~& 3P &       & 9.042 & 2.338 & 1.028 & 0.394 ~& 4D &       & 3.663 & 0.748 & 0.199 ~& 5F &  & 0.917     \\ 
4S &       &       &       & 4.820 & 1.273 & 0.531 & 0.156 ~&    &       & 11.17 & 3.130 & 1.401 & 0.571 ~&    &       & 3.814 & 0.791 & 0.210 ~&    &  & 0.929    \\ 
5S &       &       &       &       & 2.110 & 0.459 & 0.130 ~& 4P &       &       & 4.000 & 0.947 & 0.346 ~& 5D &       &       & 1.467 & 0.202 ~&    &  &      \\
6S &       &       &       &       &       & 0.883 & 0.125 ~&    &       &       & 4.709 & 1.186 & 0.463 ~&    &       &       & 1.519 & 0.209 ~&    &  &      \\
7S &       &       &       &       &       &       & 0.283 ~& 5P &       &       &       & 1.745 & 0.326 ~& 6D &       &       &       & 0.435 ~&    &  &      \\
   &       &       &       &       &       &       &       &    &       &       &       & 1.999 & 0.407 ~&    &       &       &       & 0.444 ~&    &  &      \\
   &       &       &       &       &       &       &       & 6P &       &       &       &       & 0.675 ~&    &       &       &       &  &    &  &      \\
   &       &       &       &       &       &       &       &    &       &       &       &       & 0.787 ~&    &       &       &       &  &    &  &       \\
   
\hline
$\langle \varkappa |r^{-2}|\bar{\varkappa}\rangle$ & 1S~~  & 2S~~  & 3S~~  & 4S~~ & 5S~~ & 6S~~ & 7S~~ &  &  2P~~  & 3P~~ & 4P~~ & 5P~~ & 6P~~ &  & 3D~~ & 4D~~ & 5D~~ & 6D~~ & & 4F~~ & 5F~~      \\ 
\hline
1S & 40946.& 15344.& 7382. & 3819. & 1889. & 835.5 & 247.6 ~& 2P & 712.5 & 273.7 & 134.7 & 63.35 & 24.61 ~& 3D & 88.84 & 31.07 & 12.61 & 3.519 ~& 4F &~13.36 & 2.685     \\ 
2S &       & 6502. & 3072. & 1586. & 783.9 & 346.7 & 102.7 ~&    & 2077. & 915.8 & 462.1 & 220.4 & 90.99 ~&    & 100.4 & 35.99 & 14.72 & 4.088 ~&    &~13.89 & 2.803    \\ 
3S &       &       & 1536. & 782.5 & 386.3 & 170.8 & 50.62 ~& 3P &       & 177.6 & 79.58 & 37.05 & 14.36 ~& 4D &       & 23.45 & 8.298 & 2.290 ~& 5F &  & 1.465     \\ 
4S &       &       &       & 415.2 & 202.7 & 89.52 & 26.53 ~&    &       & 508.4 & 246.3 & 116.9 & 48.24 ~&    &       & 26.45 & 9.512 & 2.618 ~&    &  & 1.517    \\ 
5S &       &       &       &       & 102.5 & 44.82 & 13.27 ~& 4P &       &       & 48.80 & 21.09 & 8.119 ~& 5D &       &       & 4.921 & 1.209 ~&    &  &      \\
6S &       &       &       &       &       & 20.31 & 5.951 ~&    &       &       & 136.2 & 62.64 & 25.76 ~&    &       &       & 5.534 & 1.365 ~&    &  &      \\
7S &       &       &       &       &       &       & 1.829 ~& 5P &       &       &       & 11.74 & 4.225 ~& 6D &       &       &       & 0.497 ~&    &  &      \\
   &       &       &       &       &       &       &       &    &       &       &       & 32.11 & 12.80 ~&    &       &       &       & 0.546 ~&    &  &      \\
   &       &       &       &       &       &       &       & 6P &       &       &       &       & 1.962 ~&    &       &       &       &  &    &  &      \\
   &       &       &       &       &       &       &       &    &       &       &       &       & 5.693 ~&    &       &       &       &  &    &  &       \\

\end{tabular}\\
\hline
\hline
\end{tabular}
\end{table}

\section{Numerical results}
\renewcommand{\theequation}{III.\arabic{equation}}
\setcounter{equation}{0}

To calculate the variance with exchange effects, Eq.~(\ref{D4F}) requires knowledge of the 
matrix elements of $r^{-1}$ and $r^{-2}$. The average values of $r^{-1}$ in the inner and outer electron orbits are approximately in the ratio of $Z:1$. The average values of $r^{-2}$ correlate roughly as $Z^2:1$. When an atom has multiple partially filled shells, Eq.~(\ref{D4F}) can be used to provide an approximate estimate. Electrons on the inner completely filled shells contribute the most to the variance, hence the accuracy of the estimate is expected to be reasonable.

Wave functions of stationary states are determined up to a phase multiplier and usually selected as real. 
The off-diagonal matrix elements are then defined up to the sign. 
The signs of radial wave functions differ in the RHF method and in the program package 
G\textsc{rasp}2018. The variance depends on squares of the off-diagonal matrix elements, so the sign arbitrariness 
has no effect on the results.


The RHF method is an excellent tool for solving problems involving the construction of atomic shells. Reference 
\cite{Clementi:1974} presents a systematic collection of the atomic wave functions through decomposition over a finite set of basis 
functions. The number of terms in the finite series and the parameters of basis functions are determined independently for each 
shell. The present paper uses these techniques to evaluate the matrix elements in atoms of the interest. The calculation results are placed in Tables \ref{tab:table20} -- \ref{tab:table52}, where calculations of 
diagonal matrix elements are also compared with the results of the relativistic DHF method \cite{Desclaux:1973}. Table 
\ref{table5} displays the exchange corrections and variance. The exchange effect contributions to 
$\mathcal{D}$ can be estimated at about 10\%.

In all examples studied, the square of variance is typically an order of magnitude greater than the average value. This is likely related to the fact that in the double--$\beta$ decay, the predominant contribution to the variance is made by rare excitations of electrons in the inner orbits. Feinberg \cite{Feinberg:1941} and Migdal \cite{Migdal:1941} calculate the probability of atom ionization due to shaking caused by knocking an electron from the K shell into the continuum.

The non-relativistic theory has a degeneracy in the total angular momentum $j = l \pm 1/2$. For testing purposes, 
we evaluated the orthonormality of the wave functions of the orbitals against the radial quantum number for each 
pair $(jl)$ and compared them to the known diagonal matrix elements from Ref.~\cite{Desclaux:1973}. In the 
relativistic models, the wave functions of orbitals $j = l \pm 1/2$ split, resulting in two 
values for each $l \geq 1$. The upper row represents $j = l + 1/2$. The orthonormality holds with an 
accuracy of at least $10^{-5}$. The agreement between the diagonal matrix elements 
is satisfactory. Deviations increase systematically with increasing $Z$, owing to relativistic effects. 
The non-relativistic approach underestimates negative moments of radius because in the relativistic 
framework the probability density of atomic electrons is shifted 
to shorter distances.
Since the off-diagonal matrix elements are 
symmetric under the interchange $n \leftrightarrow \bar{n}$, the tables display only the values $n \leq \bar{n}$. Table 
\ref{table5} contains the variance data for atoms with atomic numbers $Z\leq 54$. The constraint on $Z$ 
is due to the limitations of the non-relativistic approximation.

For the relativistic calculation of the quantities entering Eqs.~(II.10) and (II.23), the package G\textsc{rasp}2018 is used. G\textsc{rasp}2018 is based on the multi-configuration DHF approach to solve the most significant atomic 
physics problems: identifying energy levels, calculation of radiation decay rates, hyperfine structure, etc. 
The package allows to compute radial wave 
functions for any given electron configuration with definite total angular momentum. For each atom from Table \ref{table5} we 
compute a set of orbital wave functions $f_{\alpha}(r)$ and $g_{\alpha}(r)$, total energies, and matrix elements $\langle \varkappa |r^{-1}|\bar{\varkappa}\rangle$ and 
$\langle \varkappa |r^{-2}|\bar{\varkappa}\rangle$ needed to calculate the parameters $\mathcal{C}$ and $\mathcal{D}$. The values of $\mathcal{C}$ are in reasonable agreement with the results of Ref. \cite{Krivoruchenko:2023b}. 
The results for
$\Delta\mathcal{D}_{\mathrm{DHF}}^{1/2}$ and $\mathcal{D}_{\mathrm{DHF}}^{1/2}$ are compared with the non-relativistic counterparts 
$\Delta\mathcal{D}_{\mathrm{RHF}}^{1/2}$ and $\mathcal{D}_{\mathrm{RHF}}^{1/2}$. In Table \ref{tab:table60} we provide the matrix elements 
for neodymium and uranium atoms obtained with the DHF method. In this case, the splitting of the levels with $j=l+1/2$ and $j=l-1/2$ 
is significant and relativistic calculation is mandatory. 

Knowledge of the parameters $\mathcal{C}$ and $\mathcal{D}$ allows to
model the probability density of the kinetic energy of $\beta $
electrons. A popular phenomenological distribution in the finite interval 
$[0,1]$ is the beta--distribution \cite{Korolyuk:1985}. We use the beta--distribution 
to describe the probability density in the
process that occurs with the excitation of the shell. The full distribution
also takes into account the singular part associated with the probability of
the shell remaining in an unexcited state \cite{Krivoruchenko:2023b}: 
\begin{equation}
w(x)=K_{Z}^{2}\delta (x)+(1-K_{Z}^{2})\frac{\Gamma (a)\Gamma (b)}{\Gamma
(a+b)}x^{a-1}(1-x)^{b-1}.
\end{equation}
The probability is normalized by one, with $a>0$ and $b>0$. In the given
instance, $x=\epsilon /Q^{\ast }$, where $\epsilon $ is the excitation energy
of the electron shell; $Q^{\ast }=Q-I_{2}$, where $I_{2}$ is the sum of the first two ionization
potentials of the daughter atom; $K_{Z}$ is the overlap integral of the
electron shells of the parent and daughter atoms. 
The kinetic energy of $\beta $ electrons is
the difference between the magnitude of $Q^{\ast }$ and the
excitation energy of the electron shell. Since the average excitation energy
and variance are known, the parameters $a$ and $b$ can be determined from
the equations
\begin{eqnarray}
\int_0^1 x w(x)dx  &=&\frac{\mathcal{C}}{Q^{\ast }}, \\
\int_0^1 x^2 w(x)dx   &=& \frac{\mathcal{D+C}^{2}}{Q^{\ast 2}}.
\end{eqnarray}
This method for evaluating distribution parameters allows for the determination of probability distribution 
of the excitation energy. In the decay of germanium, e.g., the excitation energy does not exceed
$0.55$ keV with 95\% probability. 

\section{Conclusion}

This work finds the contribution of exchange effects to the variance of the excitation energy of the shell of 
a daughter atom in $0\nu2\beta$--decay. The effect is described by Eq.~(\ref{D4F}), which is derived in Sect.~II.B. To compute the variance, one needs to know the 
matrix elements of the values $r^{-1}$ and $r^{-2}$. These matrix elements were determined for 
eleven atoms that exhibit $2\nu2\beta$--decay. Tables I--VI provide a summary of the results. Table 
\ref{table5} shows the values of the corrections for exchange effects. The calculations utilized 
the non-relativistic RHF approach, which is justified for $Z\lesssim 54$. The results were 
compared to those obtained with the software package G\textsc{rasp}2018, which uses the DHF relativistic 
technique. Both methods are in reasonable agreement at $Z\lesssim 54$. The relativistic 
DHF approach is also used to evaluate the average excitation energy of atoms listed in Table I and, 
in particular, the impact of exchange effects in heavy 
neodymium and uranium atoms. 

The obtained results can 
be utilized to describe the spectrum of $\beta$ electrons in $0\nu2\beta$--decay, taking into consideration 
the excitation of the electron shell of the daughter atom. 

\section*{Acknowledgements}
The authors gratefully acknowledge F.F. Karpehsin for discussions. 




\end{document}